Mehta Dalip Singh (Orcid ID: 0000-0002-1569-1647)

FULL ARTICLE

# High-resolution single-shot phase-shifting interference microscopy using deep neural network for quantitative phase imaging of biological samples

Sunil Bhatt [#] | Ankit Butola [#] | Sheetal Raosaheb Kanade | Anand Kumar | Dalip Singh Mehta *

Bio-photonics and Green-photonics Laboratory, Department of Physics, Indian Institute of Technology Delhi, Hauz-Khas, New Delhi 110016, India.

# These authors contributed equally to this work.

*Correspondence

Dalip Singh Mehta, Bio-photonics and Green Photonics Laboratory, Department of Physics, Indian Institute of Technology Delhi, Hauz-Khas, New Delhi 110016, India.

Email: mehtads@physics.iitd.ac.in

Email: sunilbhatt.619@gmail.com

White light phase-shifting interference microscopy (WL-PSIM) is a prominent technique for high-resolution quantitative phase imaging (QPI) of industrial and biological specimens. However, multiple interferograms with accurate phase-shifts are essentially required in WL-PSIM for measuring the accurate phase of the object. Here, we present single-shot phase-shifting interferometric techniques for accurate phase measurement using filtered white light ($520 \pm 36$nm) phase-shifting interference microscopy (F-WL-PSIM) and deep neural network (DNN). The methods are incorporated by training the DNN to generate 1) four phase-shifted frames and 2) direct phase from a single interferogram. The training of network is performed on two different samples i.e., optical waveguide and MG63 osteosarcoma cells. Further, performance of F-WL-PSIM+DNN framework is validated by comparing the phase map extracted from network generated and experimentally recorded interferograms. The current approach can further strengthen QPI techniques for high-resolution phase recovery using a single frame for different biomedical applications.

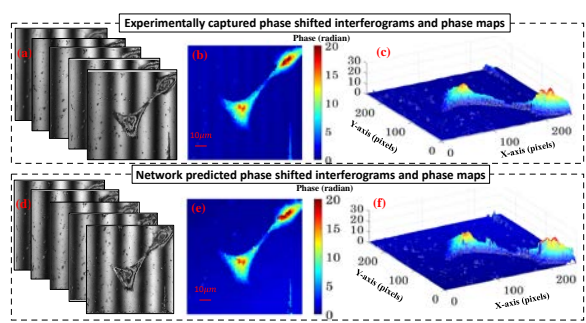

## INTRODUCTION

Quantitative phase imaging (QPI) is a rapidly emerging label-free technique to reconstruct quantitative information related to the refractive index and local thickness of the specimens[1]. The experimental and computational advancement in QPI is being widely adopted for extracting quantitative information of various industrial and biological specimens such as an optical waveguide, stem cells, human red blood cells (RBC), tissue sections, sperm samples, and among others[2-5]. In the past few decades, various newly developed QPI techniques have been implemented to improve the spatial resolution, space-bandwidth, temporal phase sensitivity, acquisition rate, and spatial phase sensitivity of the system[1, 6-8]. In QPI system, the spatial phase sensitivity and data acquisition rate are inversely related to each other. For example, coherent light-based (such as laser-based) QPI techniques offer single-shot data acquisition and phase reconstruction of the specimens but suffers with poor spatial phase sensitivity due to high spatial and temporal coherence. Single-shot phase reconstruction is possible in laser-based QPI by acquiring a high fringe density interferogram and then applying the standard Fourier transform (FT) approach[5, 8, 9]. Since the implementation of the FT algorithm requires high fringe density to separate the DC and twin images for noise-free phase recovery, so sufficiently large tilt angle is needed between the reference and object beams which essentially makes it off-axis interferometry[5].

Although fast data acquisition rate in coherent QPI technique is very popular for live-cell imaging but the presence of speckle noise degrades the spatial phase sensitivity hence unable to offer the phase map of thin biological specimens[10, 11] of thickness less than 200 nm[12]. Besides, phase reconstruction of step height larger than half the wavelength of the light cannot be extracted in a single laser-based QPI due to $2\pi$ ambiguities [13, 14]. To overcome these limitations, white light phase-shifting interference microscopy (WL-PSIM) is the most commonly used technique for ultra-sensitive measurement in both industrial and biological specimens[15-17]. WL-PSIM is based on the principle of low coherence interferometry where back-reflected light from the sample and reference arm are superimposed and forms an interference pattern only if the optical path difference (OPD) between these two arms lies within the coherence length ($L_c$) of the light source i.e., OPD $\leq L_c$. Since the coherence length of white light lies between 1-2μm, it is not possible to introduce a large tilt between reference and sample arm to get high fringe density throughout the field of view (FOV). However, introducing a constant phase shift between multiple low fringe density interferograms, it is possible to calculate the phase map of the object, and the popular approach is known as phase-shifting interferometry (PSI) [15, 18, 19].

In the past few years, a significant amount of work has been done in the development of various PSI techniques[20, 21]. In PSI, a pre-calibrated piezoelectric transducer (PZT) is used and sufficient time is required to record precise phase-shifted multiple interferograms[8]. Although PSI has the advantage of utilizing the full resolution of the system, requirement of multiple frames is the key obstacle in PSI for many applications such as live-cell imaging and measurement with dynamic samples[22]. To overcome these limitations, various single-shot phase-shifting approaches have been developed in the recent past[13, 23-26]. However, these approaches either suffer with complex experimental setups while using multiple charge couple device (CCD) cameras or significantly increases the cost of the system[17, 26, 27]. On the other hand, polarization-based PSI leads to inefficient utilization of the CCD chip[25, 27] since a micro polarizer array that provides polarized imaging with phase shift is attached to the image sensor. Therefore, a technique that can provide single-shot high-resolution phase imaging without compromising with the system resolution would be more applicable in live-cell imaging and other industrial objects.

Here, we propose single-shot phase-shifting interference microscopy (PSIM) with a low-coherent, filtered white light source ($520 \pm 36$ nm) assisted with a deep neural network (DNN) to achieve high-resolution phase imaging of both industrial and biological samples. We propose two different approaches for single-shot WL-PSIM. The first approach is incorporated by generating four phase-shifted interferograms with a single input interferogram using DNN and then using a traditional solver (five-step phase-shifting algorithm) for the phase reconstruction. The second approach is to train a DNN end-to-end to reconstruct the phase map from a single input interferogram without any intermediate phase-shifted frames. Our experimental results on the optical waveguide and MG 63 osteosarcoma cells showed the potential of the proposed framework for single-shot high-resolution phase imaging of industrial and biological samples. In addition, phase map retrieved from experimentally recorded interferograms on optical waveguide and MG 63 osteosarcoma cells is compared with phase map extracted from network-generated interferograms (in first approach) and with direct network generated phase (in second approach). The proposed framework shows its applicability for both industrial and biological samples.

# THE PROPOSED FRAMEWORK: EXPERIMENTAL SETUP, AND NETWORK ANALYSIS

## Experimental setup and five phase-shifting algorithm

The experimental setup of filtered white light ($520 \pm 36$ nm) phase-shifting interference microscopy (F-WL-PSIM) to realize single-shot high-resolution QPI is shown in Fig. 1.

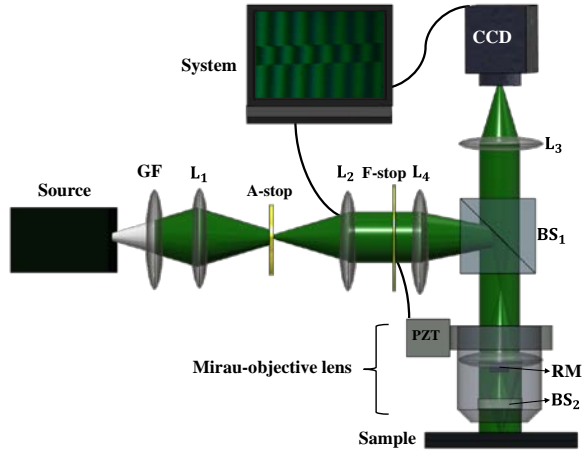

**FIGURE 1.** The schematic diagram of filtered white light ($520 \pm 36$ nm) phase-shifting interference microscope (F-WL-PSIM). GF: green-filter, $L_1, L_2, L_3$ and $L_4$: lenses, A-stop: aperture-stop, F-stop: field-stop, $BS_1$ and $BS_2$: beam splitter, PZT: piezo-electric transducer, RM: reference mirror, CCD: charge-coupled device.

In this system, the divergent light from the halogen lamp (12V50W, LV-LH50PC) is illuminated through a narrow band pass green filter (GF) of 520 nm peak wavelength with $\pm$ 36 nm bandwidth. Here, the width of the interferogram is limited and governed by the longitudinal spatial coherence (LSC) of the system instead of temporal coherence length of the light source since the LSC length is much smaller than the temporal coherence of the light source[28]. Lens $L_1$ collects the emergent filtered light and focus it on aperture stop (A-stop), and aperture stop controls the spatial coherence of light source by varying its size. Further, lens $L_2$ collects all converging light from output of A-stop and collimates, which finally focuses at back pupil of the Mirau objective by lens $L_4$ for uniform illumination at the sample plane and the field-stop (F-stop) controls field of view (FOV) of the system. The beam-splitter $BS_1$ directs light towards the sample plane through the Mirau-objective lens. We used reflection mode QPI system as it provides $2n/\Delta n$ higher sensitivity than transmission mode quantitative phase microscopy system[10].

The $BS_2$ in The Mirau objective lens (Model No. 503210, 50X/0.55 DI, WD 3.4 Nikon, Japan) reflects light towards the RM and transmits light towards the sample. The Mirau-objective lens is attached with the piezo-electric transducer (NV40 3CLE) to generate equal temporal phase-shifted data. The back-reflected light from the sample interfered with reflected light from RM at $BS_2$ which further focused by the lens $L_3$ at CCD camera (INFINITY2-1C). The captured phase-shifted interferograms further used for phase reconstruction of the specimen using the 5-step phase-shifting algorithm.

The idea behind the PSI is to introduce systematic phase-shifts between reference and sample fields. Different phase-shifting techniques have been proposed in the past to extract the phase $\phi(x, y, t, \lambda_j)$ of object, out of which five phase-shifted interferometry is preferred because of moderate phase error and acquisition time[19]. In PSI, the phase information $\phi(x, y, t, \lambda_j)$ of the object can be calculated from following expression[19, 29]:

$$\phi(x, y, t, \lambda_j) = \tan^{-1}\left[\frac{2\left(I_4(x, y, t, \lambda_j) - I_2(x, y, t, \lambda_j)\right)}{I_1(x, y, t, \lambda_j) - 2I_3(x, y, t, \lambda_j) + I_5(x, y, t, \lambda_j)}\right] \quad (1)$$

where $I_1(x, y, t, \lambda_j)$, $I_2(x, y, t, \lambda_j)$, $I_3(x, y, t, \lambda_j)$, $I_4(x, y, t, \lambda_j)$ and $I_5(x, y, t, \lambda_j)$ are the five shifted frames with a constant phase difference of $\delta(t)$ and j corresponds to different wavelength of the white light source. Our first approach of the study is to predict $I_2(x, y, t, \lambda_j)$, $I_3(x, y, t, \lambda_j)$, $I_4(x, y, t, \lambda_j)$ and $I_5(x, y, t, \lambda_j)$ from the first interferogram i.e., $I_1(x, y, t, \lambda_j)$ and then applying 5-step phase-shifting algorithm for the comparison of reconstructed phase map from the experimentally-recorded and network-generated interferograms. The second approach of our study is to predict direct phase $\phi(x, y, t)$ of the object from the single input interferogram $I_1(x, y, t)$. To achieve this, we used and trained a generative adversarial network (GAN).

**The architecture of Generative adversarial network (GAN)**

The GAN architecture comprised simultaneous training of two models generator

(G) and discriminator (D)[30, 31]. The generator model took an image as input dataset and generate some mapped images. The discriminator differentiate between input target images (ground truth) and generated images (output of generator). While differentiating between ground truth and generated images, the discriminator calculate the losses and give feedback to the generator for fine-tuning. The GAN architecture developed in the present study has been shown in Fig. 2. The detailed discussion about GAN architecture and mathematical equations is given in the supplementary file.

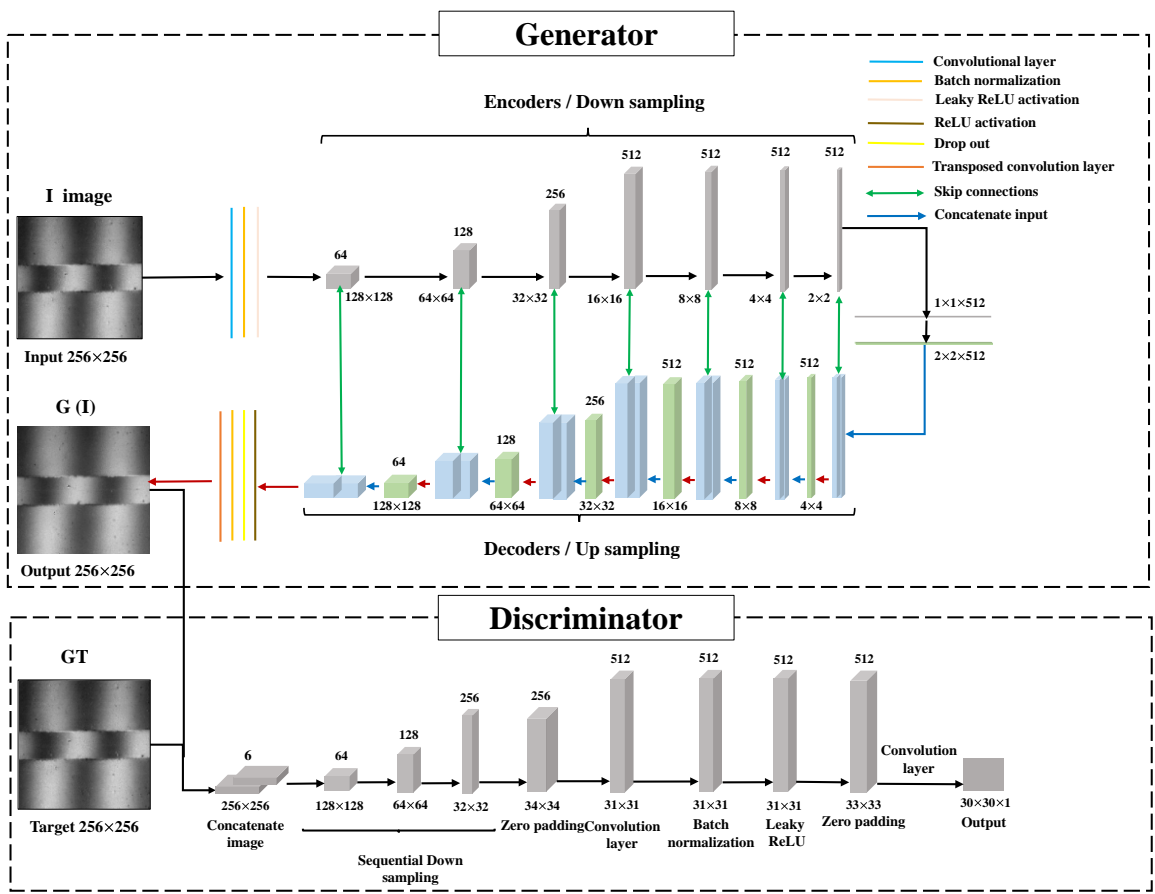

**FIGURE 2.** Generative adversarial network (GAN) architecture for single-shot high-resolution quantitative phase imaging using phase-shifting interference microscopy.

The model is trained, optimized and test on experimentally recorded datasets of optical waveguide and MG63 osteosarcoma cell. The dataset is divided into two parts: training (~80%) and test datasets (~20%). During training, network figured out the losses between the generated images and the ground truth i.e., generator and discriminator losses[30-32]. These losses were then feedbacked to the corresponding models (as shown in Fig. 3), until the discriminator achieved equal probability (i.e., ~1/2) of getting ground truth and generated images[30]. For testing we used different datasets that network has never seen during the training process.

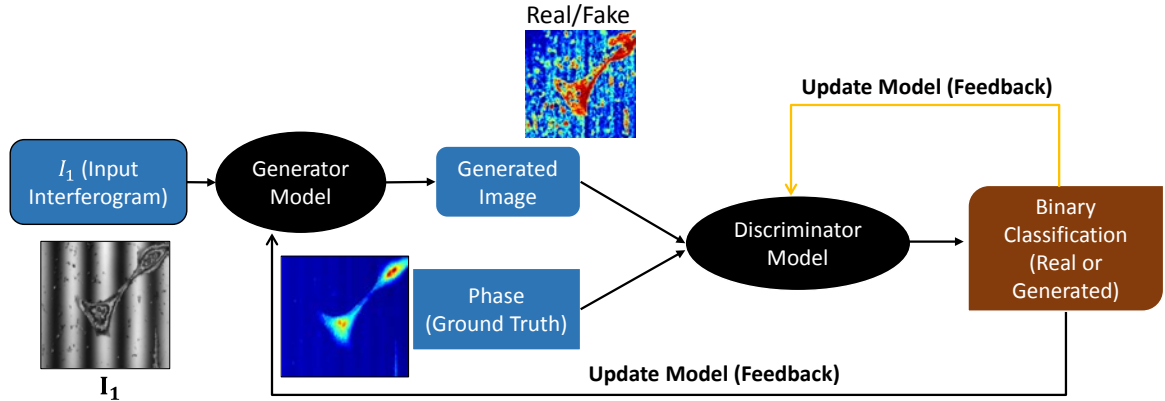

**FIGURE 3.** Block diagram of the architecture of generative adversarial network (GAN).

## RESULT AND DISCUSSION

The experimental verification of the proposed framework is shown on two different datasets i.e., optical waveguide and MG63 osteosarcoma cell for both the approaches.

### (1) Generating four phase-shifted frames with a single frame using GAN and a traditional solver for phase reconstruction:

In this approach, the training and testing of the network accomplished with a total 312 datasets of phase-shifted interferograms recorded by F-WL-PSIM system for waveguide dataset. The 312 waveguide datasets consist of 5 phase-shifted interferograms and therefore, a total of 1560 interferograms is used for training and testing purposes of the network. Similarly 240 datasets i.e., 1200 interferometric images of MG63 cells is used. The cell culture process of MG 63 cell data is provided in the supplementary file. Further, we used data augmentation process to increase the trainable parameters for both the datasets. Data augmentation is a pre-processing step for image augmentation such as rotation, resizing and reflection which results in more robust training of the network. During training of the network, first two phase-shifted frames were combined to predict phase-shift between $I_1$ and $I_2$ and to generate an image $I_2'$. Similarly $I_2 \rightarrow I_3$, $I_3 \rightarrow I_4$ and $I_4 \rightarrow I_5$ were taken as input to generator model. For testing the network, we used different datasets that network never seen during training process. The network is trained separately for both datasets in both the approaches since the characteristic features extracted from sample during training of the network may vary significantly, and the pre-trained network may be biased to previously encoded features. We optimized the hyperparameters for both waveguides to MG 63 cell datasets in both the approaches. The generator model has 54,414,979 and the discriminator model has 2,768,641 trainable parameters in the network.

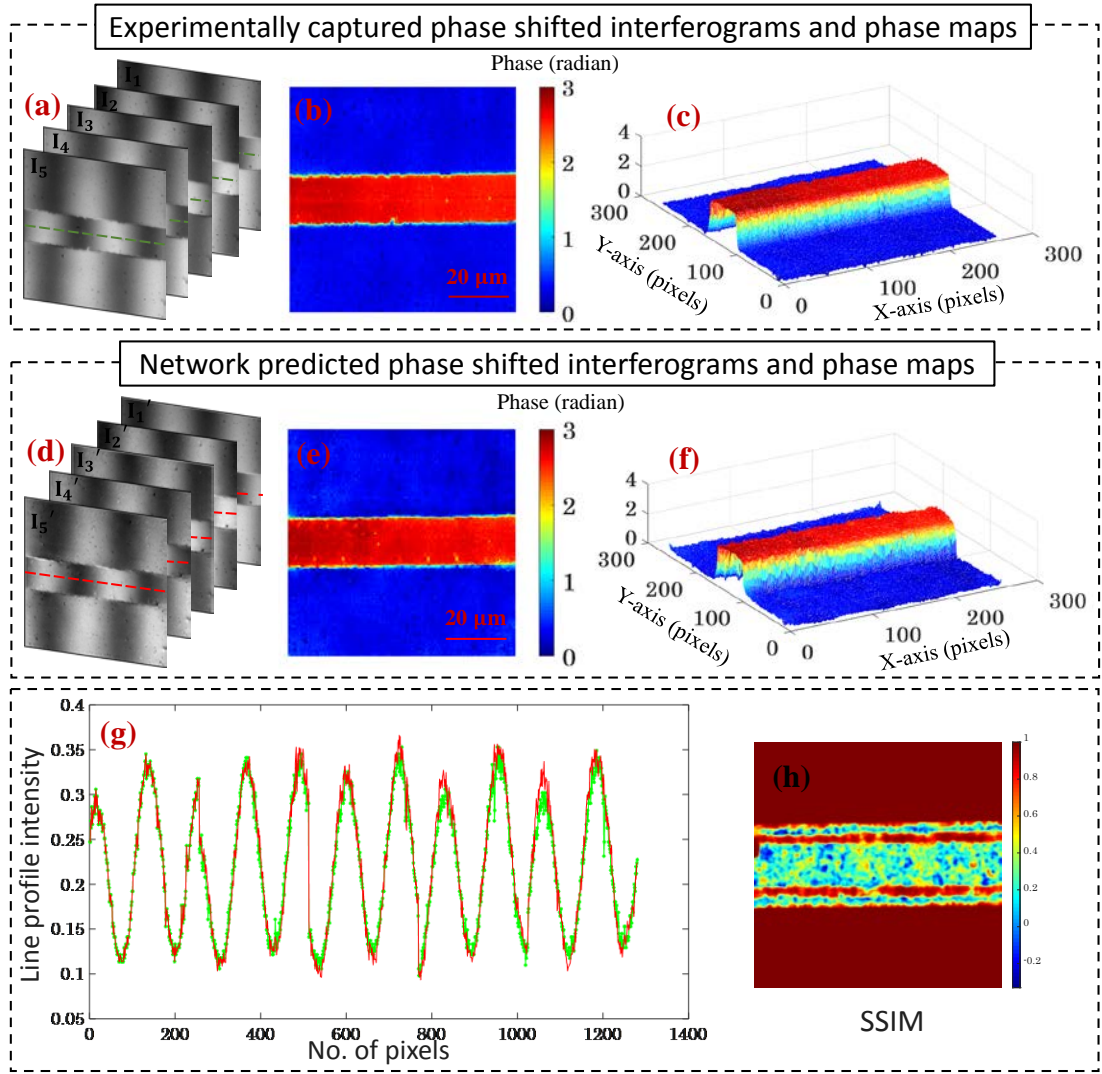

**FIGURE 4.** Comparison between experimental and network-generated five phase-shifted interferograms and their corresponding phase-maps on waveguide structure. (a) experimentally captured phase-shifted interferograms. (b) and (c) corresponding 2D, and 3D reconstructed phase-map respectively. (d) network-generated interferograms. (e) and (f) corresponding 2D, and 3D reconstructed phase-maps, respectively. (g) line profile comparison plot of equal five phase-shifted interferograms for experimentally captured (green) and network-generated (red). (h) represents the structured similarity index measure (SSIM) between experimental (ground truth) and network predicted phase. The SSIM index for our waveguide sample is 0.80.

Figure 4 depicts the comparison between reconstructed phase-maps of experimentally captured and network-generated interferograms. The network trained on first interferogram to predict four equally phase-shifted frames. We selected the first frame i.e., $I_1$ as an input of the network since the network is trained with a relative phase difference between $I_1 \rightarrow I_2, I_2 \rightarrow I_3, I_3 \rightarrow I_4$ and $I_4 \rightarrow I_5$.

Figure 4 (a) represents result obtained from multi-shot PSI with pre-calibrated PZT captured in the F-WL-PSIM setup. Figure 4(b), (c) shows 2D and 3D reconstructed phase maps of the optical waveguide using five-step phase-shifting algorithm, respectively. On the other hand, Fig. 4(d) represents network-generated interferograms and Fig. 4(e), (f) are their corresponding 2D, and 3D reconstructed phase maps, respectively.

In addition, the line profile along the centre of experimentally captured (Fig. 4(a)) and network-generated (Fig. 4(d)) interferograms is compared. The line profile is calculated to check phase-shift accuracy in predicted and experimentally captured interferograms. Note that, all five frames (image size (0,256)) have been stitched one after the other to plot line profile. The x-axis of the line profile comparison shows number of pixels i.e., first (0, 256) interval corresponds to first phase-shifted interferogram i.e. $I_1 \rightarrow I_1'$, second (256, 512) $I_2 \rightarrow I_2'$ and so on. The line profile comparison (see Fig. 4(g)) shows exact match in phase-shift between frames of network generated (red) and experimentally captured (green) five phase-shifted interferograms.

Further, the similarity in structure between output data with input can be quantified by the structural similarity index measure (SSIM). The SSIM value varies between [-1 1], where 1 represents identical predicted and ground truth images. In our case, the SSIM value was found 0.80 (see Fig. 4(h)) for waveguide phase data between Fig. 4(b) and (e). There is a slight mismatch in SSIM between the phase map calculated from experimentally captured and network predicted interferograms. The mismatch in SSIM may occur due to phase-related artifacts during the data acquisition[8] such as spatial phase sensitivity, temporal phase sensitivity, and mismatch of equal phase-shift between the data frames.

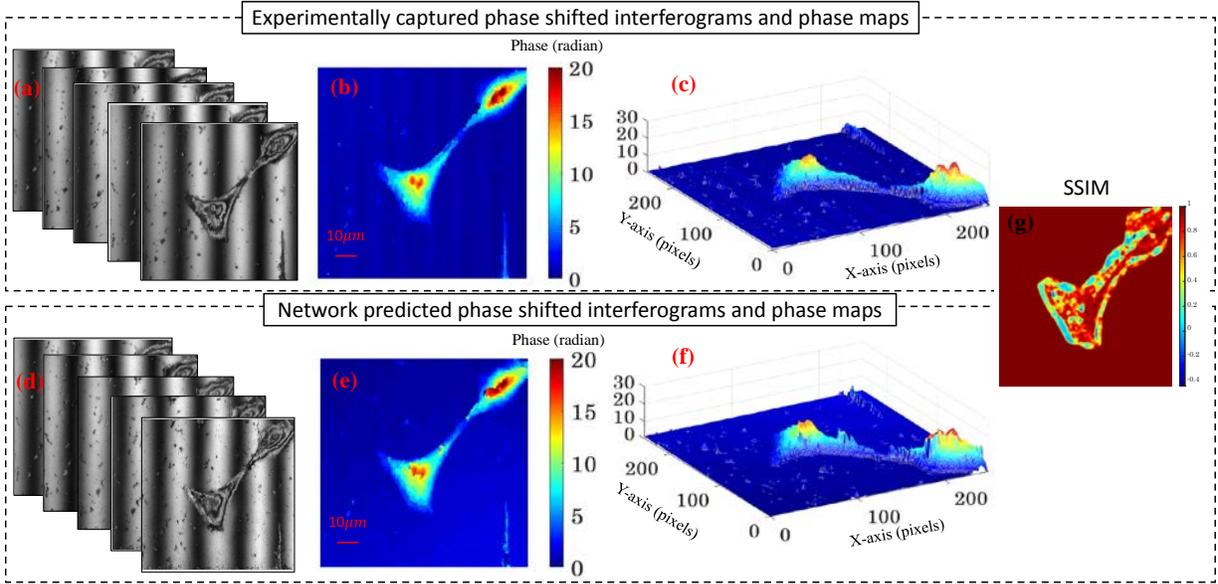

**FIGURE 5.** Comparison between the experimental and network generated five phase-shifted interferograms on MG63 osteosarcoma cell. (a) experimentally captured phase-shifted interferograms. (b) and (c) corresponding 2D and 3D reconstructed phase-map. (d) network-generated phase-shifted interferograms. (e) and (f) corresponding 2D, and 3D reconstructed phase-map. (g) structured similarity index measure (SSIM) between experimental (ground truth) and network predicted phase.

Figure 5 depicts comparison between reconstructed phase maps from experimentally recorded and network generated interferograms of MG63 osteosarcoma cells. Figure 5(a) represents the result obtained from multi-shot PSI with well-calibrated PZT. Figure 5(b) and (c) shows corresponding 2D and 3D reconstructed phase maps respectively. Fig. 5(d) represents network-generated phase-shifted interferograms and Fig. 5(e), (f) are corresponding 2D and 3D reconstructed phase maps, respectively. Figure 5(g) shows the SSIM between the experimental (ground truth) and network-generated phase. The SSIM value was found 0.92 (see Fig. 5(g)) for MG63 cell phase data between Fig. 5(b) and (e). However, slight mismatch between experimental and network predicted phase can be seen in the SSIM and cannot be avoid in practice due to phase sensitivity, diversity in the datasets, temporal stability of the system and limitation of the network. Furthermore, the background modulations occur because of unequal phase-shift due to environmental fluctuations between experimentally captured five phase-shifted interferograms. Whereas, background modulation is completely disappeared in network predicted phase and possible reason might be averaging the error between phase-shift due to large training of the network which can be considered as another advantage of the proposed framework.

**(2) Training GAN end-to-end to generate the phase map from a single frame:**

In our second approach, we used similar datasets of optical waveguide and MG 63 cell data for training and testing of the network. For training and testing purpose, total 312 five phase-shifted data sets of optical waveguide and total 240 five phase-shifted data sets of MG 63 cell data were recorded using F-WL-PSIM setup. We have reconstructed corresponding phase map from the recorded datasets by applying five-step phase-shifting algorithm. We extract the first interferogram of each sample dataset (i.e., $I_1$ from each dataset) for training and testing of the network. Total 270 single interferogram of optical waveguide and 210 single interferogram of MG 63 cell datasets is used for training and rest of the single data for the testing purpose. Further, data augmentation process with rotation of 30 degree increase the trainable parameters for both the datasets. Thus, finally at the time of training we have total 3,240 augmented data of interferogram as well as phase for optical waveguide and 2,520 augmented data for MG 63 cell structure.

During training of the network, the single interferogram ($I_1$) and its reconstructed phase (Ground truth) is combined input to the network for training. After sufficient training, the network generate phase image similar to the ground truth with minimum loss function. Thus, after sufficient amount of training network provides direct phase image for the single input test interferogram.

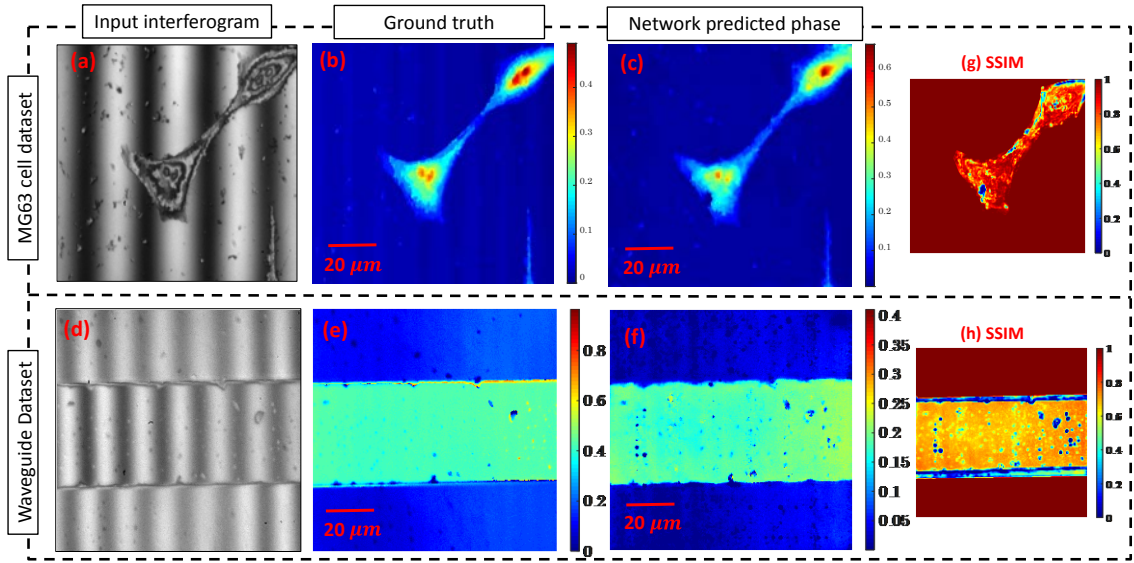

**FIGURE 6.** Comparison between ground truth and network predicted phase image on MG63 osteosarcoma cell and optical waveguide dataset. (a), (d) are the experimentally captured interferogram of MG 63 cell and waveguide data, respectively. (b), (e) are the corresponding 2D reconstructed phase-map using five-step phase-shifting algorithm, respectively (c), (f) are the network-generated phase map from a single input interferogram of MG 63 cell and waveguide data, respectively. (g) structured similarity index measure (SSIM) between ground truth (b) and network predicted phase (c) of MG 63 cell data. (h) SSIM between ground truth (e) and network predicted phase (f) of waveguide data.

Mehta Dalip Singh (Orcid ID: 0000-0002-1569-1647)

Figure 6 depicts the comparison between reconstructed phase maps from experimentally-recorded five phase-shifted interferograms and network-generated phase map. Figure 6(a-c) depicts input interferogram ($I_1$), 2D reconstructed phase map using five-step phase-shifting algorithm and predicted phase by the trained network for MG 63 datasets, respectively. Similarly, Fig. 6 (d-f) shows input interferogram, 2D phase map reconstructed using phase-shifting algorithm and predicted phase by the trained network for waveguide structure, respectively. Figure 6(g) and (h) shows the SSIM between the ground truth and network-generated phase map. The SSIM value was found to be 0.95 (see Fig. 6(g)) for MG63 cell phase data and 0.80 for optical waveguide dataset.

However, careful examination must be required to interpret the F-WL-PSIM+DNN framework, although the average SSIM might decrease for foreground region only which is obvious and cannot be avoided due to many experimental and computational reasons. The experimental inaccuracy such as effect of environmental fluctuations in phase-shifting data acquisition, sample preparations, mechanical stability, spatial and temporal phase sensitivity of the system certainly affect performance of the proposed framework. On the other hand, optimization of the network such as hyper parameter tuning, batch-size, epoch-size, number of convolutional layers and size of the datasets will also affect the outcome of the framework.

## CONCLUSION

We have demonstrated F-WL-PSIM + two different DNN framework for single-shot high-resolution phase imaging. The first proposed framework generates four equally phase-shifted interferograms from one single input interferogram by using deep learning architecture. The current framework was trained and validated with experimentally captured interferograms of the optical waveguide and MG63 osteosarcoma cell.

The second proposed framework generates direct phase from single interferogram. The proposed framework does not require any intermediate multiple phase-shifted data frames and phase-shifting algorithm for the phase reconstruction. This approach can provide direct phase map for a single interferogram from the F-WL-PSIM, which excludes the use of well calibrated PZT for the data acquisition.

Moreover, networks-predicted and experimentally-reconstructed phase map comparison and calculated SSIM value quantifies that both frameworks provide similar results for the same datasets. Therefore, one can apply any of the proposed framework as per their requirements.

Moreover, our proposed method shows high-resolution quantitative phase microscopy (QPM) using a single interferogram which negates the requirement of traditionally recorded multiple interferograms. Hence, the data acquisition time can be reduced at least by a factor of five. Furthermore, no mechanical moving phase-shifting is required, so time reduction can be more than 5 fold. The present methods are useful for the QPI of biological cells and tissues such as human RBCs, and cancer tissues where high-resolution phase imaging is important in disease diagnosis and identification of cancer margin.

# ACKNOWLEDGMENT

SB would like to acknowledge CSIR, India for financial support for this research work through the research fellowship. The author would like to acknowledge Supriya Mehta and Neetu Singh, 'Centre for Biomedical Engineering' for providing MG63 osteosarcoma cells for experimentation.

# AUTHOR CONTRIBUTIONS

F-WL-PSIM+DNN system is developed by SB and AB. SB, AB, SRK, and AK acquired the waveguide and MG-63 osteosarcoma cell datasets. The training parameters and performance evaluation of GAN architecture are optimized by SB and SRK. SB and AB prepared the first draft of the manuscript and all authors contributed towards the writing of the manuscript. SB and AB contributed equally to this work. The work is supervised by DSM.

**CONFLICT OF INTEREST**

The author declares that there are no conflicts of interest.

**DATA AVAILABILITY STATEMENT**

Data available on request from the authors.